\documentclass{emulateapj}

\usepackage{apjfonts}

\usepackage{graphics}

\input psfig

\input epsf

\lefthead{Mosquera Cuesta et al. }

\righthead{GRB HD and energy conditions}

\newcommand{\be}{\begin{equation}}

\newcommand{\ee}{\end{equation}}

\begin{document}

\pretolerance=10000

\slugcomment{The Herman's Team }

\shorttitle{HD of GRBs and prevalence of energy conditions in cosmology}

\shortauthors{Mosquera Cuesta et al.}




\title{Hubble diagram of gamma-ray bursts and prevalence of 
energy conditions in Friedmann-Lema\^itre-Robertson-Walker Cosmology}

\author{Herman J. Mosquera Cuesta\altaffilmark{1,2}, Carlos A. Bonilla 
Quintero\altaffilmark{1}, Habib Dumet M. \altaffilmark{1}, Cristina 
Furlanetto\altaffilmark{1}, Jefferson Morais\altaffilmark{1} and Rodrigo 
Turcati\altaffilmark{1} }

\altaffiltext{1}{Instituto de Cosmologia, Relatividade e Astrof\'{\i}sica
(ICRA-BR), Centro Brasileiro de Pesquisas F\'{\i}sicas \\ Rua Dr. Xavier 
Sigaud 150, Cep 22290-180, Urca, Rio de Janeiro, RJ, Brazil --- e-mail: 
hermanjc@cbpf.br }

\altaffiltext{2}{Abdus Salam International Centre for Theoretical Physics,
Strada Costiera 11, Miramare 34014, Trieste, Italy}





\date{\today}

\begin{abstract}
Since the introduction of the {\sl Amati relation} (Amati et al. 2002), other 
largely promising techniques have appeared in an attempt to turn GRBs into reliable 
cosmological probes. These incluide the {\sl Ghirlanda relation} (Ghirlanda et al. 
2004) and the recently discovered {\sl Firmani et al. relation} (Firmani et al. 
2006a),  all of which take into account the most relevant physical properties of 
gamma-ray bursts (GRBs) as peak energy, jet openning angle, and both time lag and 
variability. Such discoveries hints at the long-sought {\sl Holy Grail} of creating 
a cosmic ruler from  GRBs observables to be achievable. In this {\sl Letter} we 
construct the Hubble diagram (HD) for the standard Friedmann-Lema\^itre-Robertson-Walker 
(FLRW) cosmological model after enforcing it with the general relativistic energy 
conditions, heeding to investigate whether it still stands on as the leading scenario 
for cosmology in face of the distance modulus-redshift relation of a sample of GRBs 
that had their redshifts properly estimated and corrected upon applying on the data 
analysis the above quoted relations. Our $\chi^2$ analysis support the view that FLRW 
plus the strong energy condition (SEC) is what better fits the GRB data. But this is 
not the whole story, since for a cosmological constant $\Lambda \neq 0$ the FLRW+SEC 
analysis with undefined $p=p(\rho)$ suggests that $\Lambda$ is not constant anyhow, 
because it does not follows the $p= \omega \rho$ HD, with $\omega=-1$. Thenceforth, 
one concludes that either the cosmological constant does not exist at all, and 
consequently one is left with the FLRW+SEC which fits properly the GRBs HD and there 
is no late-time acceleration, or it does exist but should be time-varying. This all is 
contrary to current views based on SNIa observations that advogate for an actual 
constant $\Lambda$ and an accelerating universe. In connection to the results above, 
this last argumentation would imply that either there is something wrong with the SNIa 
interpretation regarding cosmic late-time acceleration (Middleditch 2006) or we will 
have to move away from general relativity. We would better to think that general 
relativity is still the most correct theory of gravity.


\end{abstract}



\section{ Introduction: GRB standardized candles for accurate cosmology }

Gamma-ray bursts (GRBs) are the biggest explosions in the universe. The major breakthrough 
in understanding of GRBs came
with the Beppo-SAX satellite discovery of  the X-ray afterglow of  GRB970228 (Costa et al. 
1997). A swift follow-up of this event allows the subsequent detection of residual optical 
and radio emissions from this transient (van Paradijs et al. 1997; Frail et al. 1997). 
From these afterglows the first measurement of the redshift of a GRB was done, what 
definitely confirmed a long-standing suspicion that GRBs have cosmological origin. Such 
realization opens-up the possibility of using GRBs for a couple of fundamental researches 
in contemporary astrophysics and cosmology: a) as dust absorption-free 
tracers of the massive star formation history in the universe (Totani 1997; Pacsynski 1998; 
Wijers et al. 1998; Blain \& Natarajan 2000, Lamb \& Reichart 2000; Bromm \& Loeb 2002, 
Lloyd-Ronning \& Ramirez Ruiz 2002; Lloyd-Ronning, Fryer \& Ramirez Ruiz 2002; Firmani 
et al. 2004; Yonetoku 
et al. 2004; Friedman \& Bloom 2005; Schaefer 2006\footnote{Regarding the recent claim by 
Schaefer (2006) on the exclusion of the cosmological constant $\Lambda$ after confronting 
it with the GRB HD, a very interesting counter-argument was provided by Friedman 
(2006). He argues that the correct HD  should also show data calculated for a 
cosmology with $\Lambda$, and that only a comparison of relative $\chi^2$ can determine the 
favored cosmology. We agree with this statement, and as we demonstrate below that condition 
should be supplemented with the need for the energy conditions to be satisfied, too.}), and 
b) as standard candles and hence as cosmic probes with which one can follow back in time upto 
very high redshifts the expansion history of the universe (Schaefer 2003; Ghirlanda et al. 
2004; Dai, Liang \& Xu 2004; Firmani et al. 2005; Liang \& Zhang 2005; Xu et al. 2005). 

Unfortunately, to accomplish both of these key investigations is not an easy task since both
face the problem of the small statistics of events with 
measured readshift ($z$) because, to determine $z$, is required deep-space optical/infrared 
or X-ray spectra. With regard to the item b), the large dispersion of the energetics of 
GRBs precludes of granting GRBs the status of cosmic rulers (Bloom, Frail \& Kulkarni 2003).
Nonetheless, a major effort is being pursued to overcome both the above mentioned drawbacks 
with significant achievements after realizing that there exist tight relations that connect 
GRB intrinsic energetics and/or luminosity with other physical observables (Ghirlanda et al. 
2004; Firmani et al. 2006a). These relations become powerful distance indicators, and in 
events such supernovae (Woosley \& Bloom 2006); for which $z$ can be inferred via an 
energetics-independent procedure, GRBs become standard candles for practical cosmographic 
studies (see Firmani et al. 2006a for a full accounting of this impressive progress). 


Presently, after the first series of controversial statements on the viability of granting 
to GRBs the status of standard cosmic rulers (Bloom 2002; Schaefer 2002, 2003; Bloom, Frail 
\& Kulkarni 2003; Friedman \& Bloom 2005), a definite consensus appears to be arising, and 
the hope to have a {\sl Holy Grail} to do cosmology upon GRBs is renascenting (Ghisellini \&
Lazzati 2004 (GGL2004); Friedman \& Bloom 2005;  M\"ortsell \& Sollerman 2005; Schaefer 2006).
GGL2004 gave to the search for a {\sl Holy Grail} a renascence after finding that the Amati et 
al. (2002) correlation becomes much tighter if one corrects $E_{\rm iso}$ for the collimation 
factor of the GRB opening angle $\theta_{\rm j}$, which is the rest-frame time of the 
achromatic break in the light curve of an afterglow.\footnote{It has been pointed out that the 
presence of a collimated break in GRB afterglow light curve is a strong evidence that the GRB 
emission is collimated into a cone of semi-aperture angle $\theta_{\rm j}$.} Thus the 
angle-corrected  $\gamma$-energetics reads: $E_{\gamma} = (1 - \cos \theta_{\rm j}) 
E_{\rm iso}$. The $E_{\gamma} - E_{\rm peak}$ correlation\footnote{This is a power-law 
relation $E_{\rm peak} - E_{\gamma}$, where : $E_{\rm peak} = \tilde{\kappa} \left( 
\frac{E_\gamma} {E_0} \right)^\eta $, that relates the energy at the maximum ($E_{\rm peak}$) 
of a given GRB with its total energy in $\gamma$-rays ($E_{\gamma}$), with the constant $E_0 
= h^{-3/2}$ yielding the best-fit parameters $\tilde{\kappa}$ and $\eta$ independent of the 
value of the Hubble parameter $h$.} have been used to put constraints on some cosmological 
parameters avoiding the {\sl circularity problem} (Ghirlanda et al. 2004; Firmani et al. 
2005). Besides, Firmani et al. (Firmani et al. 2006a) recently discovered new tigh correlation 
among prompt emission properties in long GRBs, which involves the isotropic peak luminosity, 
$L_{\rm iso}$, the peak energy, $E_{\rm peak}$, in the spectrum, $\nu L_\nu$, of the 
time-integrated prompt emission and the ``high signal" time-scale, $T_{0.45}$, that was 
previously used to characterize the variability behaviour of GRBs. In the source rest-frame, 
this relation follows as: $L_{\rm iso} = \tilde{K} E_{\rm peak}^{1.62} T_{0.45}^{ -0.49} $,
with $\tilde{K}$ a constant. It uses the GRB openning angle-corrected redshifts into $L_{\rm 
iso}$.


Here we investigate whether the energy conditions (ECs) of general relativity (GR) when 
enforced on the standard  cosmological FLRW model can bring it a pace with the Hubble 
diagram (E. P. Hubble, 1929, Proc. Nat. Acad. Sci. USA, 15, 168) constructed from the 
sample of GRBs that have redshifts measured through spectroscopy or other methods (like 
photometry). For the present analysis we shall use a set of three samples of GRBs that 
had their redshifts estimated: a) Bloom, Frail \& Kulkarni (2003; BFK2003), as shown in 
Table-1 and Figure 1a). b) The Schaefer (2006) GRB set analyzed with five methods, shown 
in Figure 1b). c) the $\theta_{\rm j}$-corrected GRBs sample, as from the discovery by 
Firmani et al. (2006a), plotted in Figure 1c).






\begin{table}
\begin{center}
\caption{Selected sample of 24 GRB events with measured redshift, computed distance 
modulus and error bars, as tabulated by Bloom, Frail \& Kulkarni (2003). \label{tbl-1} }
\begin{tabular}{ccccc}
\tableline
\tableline
GRB & Redshift & Dist.-Modulus & error & $\theta_{\rm j}$-Corrected\footnotemark[1] \\
event & $z$ &  $\mu(z)$ & $\pm$ &  $\bar{z}$\\
\tableline
970508 & 0.8349 & 45.10 & 0.47 & -\\
970828 & 0.9578 & 43.86 & 0.45 & 1.212\\
971214 & 3.4180 & $<$47.83 & 0.37 & 4.801\\
980613 & 1.0969 & $<$ 47.03 & 0.54 & - \\
980703 & 0.9662 & 44.31 & 0.45 & 1.104\\
990123 & 1.6004 & 44.00 & 0.53 & 1.358\\
990510 & 1.6187 & 47.13 & 0.42 & 1.210\\
990705 & 0.8424 & 43.98 & 0.43 & 1.109 \\
990712 & 0.4331 & $<$42.13 & 0.33 & -\\
991208 & 0.7055 & $>$43.35 & 0.86  & -\\
991216 & 1.0200 & 44.02 & 0.83 & 0.823\\
000131 & 4.5110 & $>$47.06 & 0.44 & 4.957\\
000210 & 0.8463 & $<$43.90 & 0.29 & -\\ 
000301C & 2.0335 & 46.30 & 0.40 & -\\
000418 & 1.1181 & 42.97 & 1.43 & -\\
000911 & 1.0585 & $>$44.00 & 0.46 & 1.173\\
000926 & 2.0369 & 45.93 & 0.92 & -\\
010222 & 1.4768 & 45.28 & 0.21 & -\\
010921 & 0.4509 & 41.71 & 0.62 & -\\
011121 & 0.3620 & $<$41.35 & 1.08 & -\\
011211 & 2.1400 & 47.54 & 0.45 & -\\
020405 & 0.6899 & 43.99 & 0.67 & -\\
020813 & 1.2540 & 45.00 & 1.07 & -\\
021004 & 2.3320 & 46.48 & 1.80 & -\\
\tableline
\end{tabular}
\footnotetext[1]{Redshifts in this column were corrected by using the {\sl Firmani et 
al. relation} }
\end{center}
\end{table}


\section{Energy conditions in General Relativity}

It is well-known in GR that the energy conditions (ECs) [at least seven types are currently 
invoked in GR (Visser 1996)] provide definite constraints on the physical properties of any 
matter field that could be present 
in the universe. Any degree of arbitrariness in the energy-momentum tensor $T_{\alpha 
\beta}$, or in other words; in the equation of state (EoS)\footnote{The relation between the 
energy density ($\rho$) content and the pressure ($p$).} describing a given dynamical field, 
can be thoroughly restricted by imposing on its dynamics the ECs. As a rule of pratice in
GR, such conditions are stated in a coordinate-invariant fashion by considering the timelike, 
null-like or spacelike nature of the proper $T_{\alpha \beta} = {\rm diag}(\rho, p, p, p)$ 
and any vector field $A_\gamma$ that may exist. However, it is pertinent to this discussion 
to point out that despite being physically sound, the ECs were not always universally accepted 
and issues regarding how fundamental, as a whole, the ECs are have been since long a matter of 
concern. Indeed, by the late 1970's researchers realize that not any matter field accomplish 
with the ECs. A classical outlier is the Einstein cosmological constant $\Lambda$ which, {\sl
if being indeed a constant with $\omega =-1$}, violates the strong energy condition (SEC) while 
keeps on fulfilling the WEC. Other examples are the {\sl Casimir effect} that violates the WEC 
and DEC, {\sl Hawking evaporation} which violates the NEC and {\sl cosmological anti-de Sitter 
($\rho < 0$) inflation} which violates WEC and DEC (Visser 1996). \linebreak

The ECs of relevance for the present analysis can be expressed as follows:\footnote{
We notice that recently Santos, Alcaniz \& Rebou\c cas (2006) have shown that the GR ECs 
are relevant to constraint general energy-momentum tensors (or equations of state) of current 
use in most cosmological models. It was shown that in the case of the FLRW scenario, ECs 
provide model-independent bounds on the distance-modulus of cosmic sources with dependence on 
the redshift. Data from SNIa were analyzed and the question regarding a possible violation of 
the ECs in FLRW model was addressed. They conclude that SNIa seems to violate all the ECs. 
Our analysis below in some sense come across with this conclusion. In similar lines, P\'erez 
Bergliaffa (2006) has demonstrated that the WEC is a very good tool to constraint theories 
of gravity of the kind $f(R)$.}

- Null energy condition (NEC): the quantity $T_{\alpha \beta} {\cal{N}}^\alpha 
{\cal{N}}^\beta \geq 0$ for any null vector ${\cal{N}}^\gamma \exists T_{a}(M)$ 
lying at the point $a $ in the tangent space $T(M)$ of a real four-dimensional 
manifold $M$. Physically, it means that the {\sl local} energy density as measured 
by any timelike observer has to be positive. 

- Weak energy condition (WEC): the inequality $T_{\alpha \beta} {\cal{T}^\alpha} 
{\cal{T}^\beta} \geq 0$ holds for all timelike vector ${\cal{T}^\gamma} \exists 
T_{a}(M)$. By continuity it implies the NEC bound $\rho + p \geq 0$. It follows 
that

\be
{\rm WEC}  \Longrightarrow \rho \geq 0 \hskip 0.5 truecm {\rm and} \hskip 0.2 truecm 
\rho + p \geq 0 \; .
\label{WEC}
\ee

- {Dominant energy condition (DEC): the relation} $T_{\alpha \beta} 
{\cal{T}^\alpha} {\cal{T}^\beta} \geq 0$ holds for all timelike vector ${\cal{T}^\gamma} 
\exists T_{a}(M)$ whenever it is supplemented by demanding that $T_{\alpha \beta} {\cal{T}^\alpha}$ 
be a non-spacelike vector. It states in physical terms the positivity of the {\sl local} 
energy density, and that the energy flux is time-like or null. It implies the WEC and by 
continuity the NEC. It follows that

\be
{\rm DEC}  \Longrightarrow \rho \geq 0 \hskip 0.5 truecm {\rm and} \hskip 0.2 truecm 
- \rho \leq p \leq \rho \; .
\label{DEC}
\ee

- Strong energy condition (SEC):  the quantity $(T_{\alpha \beta} - \frac{1}{2} T g_{\alpha 
\beta}) {\cal{T}^\alpha} {\cal{T}^\beta} \geq 0$, should be held for any timelike vector 
${\cal{T}^\gamma}$, being $T$ the trace of the tensor $T_{\alpha \beta}$.  It follows that

\be
{\rm SEC}  \Longrightarrow \rho + 3p \geq 0 \hskip 0.5 truecm {\rm and} \hskip 0.2 truecm 
\rho + p \geq 0 \; .
\label{SEC}
\ee

Next we impose the ECs to the FLRW model to build the HD they predict so as 
to compare it with the one constructed upon the GRB data. We will check for any putative 
violation of ECs by the observed GRBs. The best fit to the GRB $\mu(z)$ vs. $z$ relation 
(HD) will select the cosmological model one must seriously should take for 
cosmography studies.


\begin{figure*}
\centering{
\includegraphics[height=1.5in,width=1.7in]{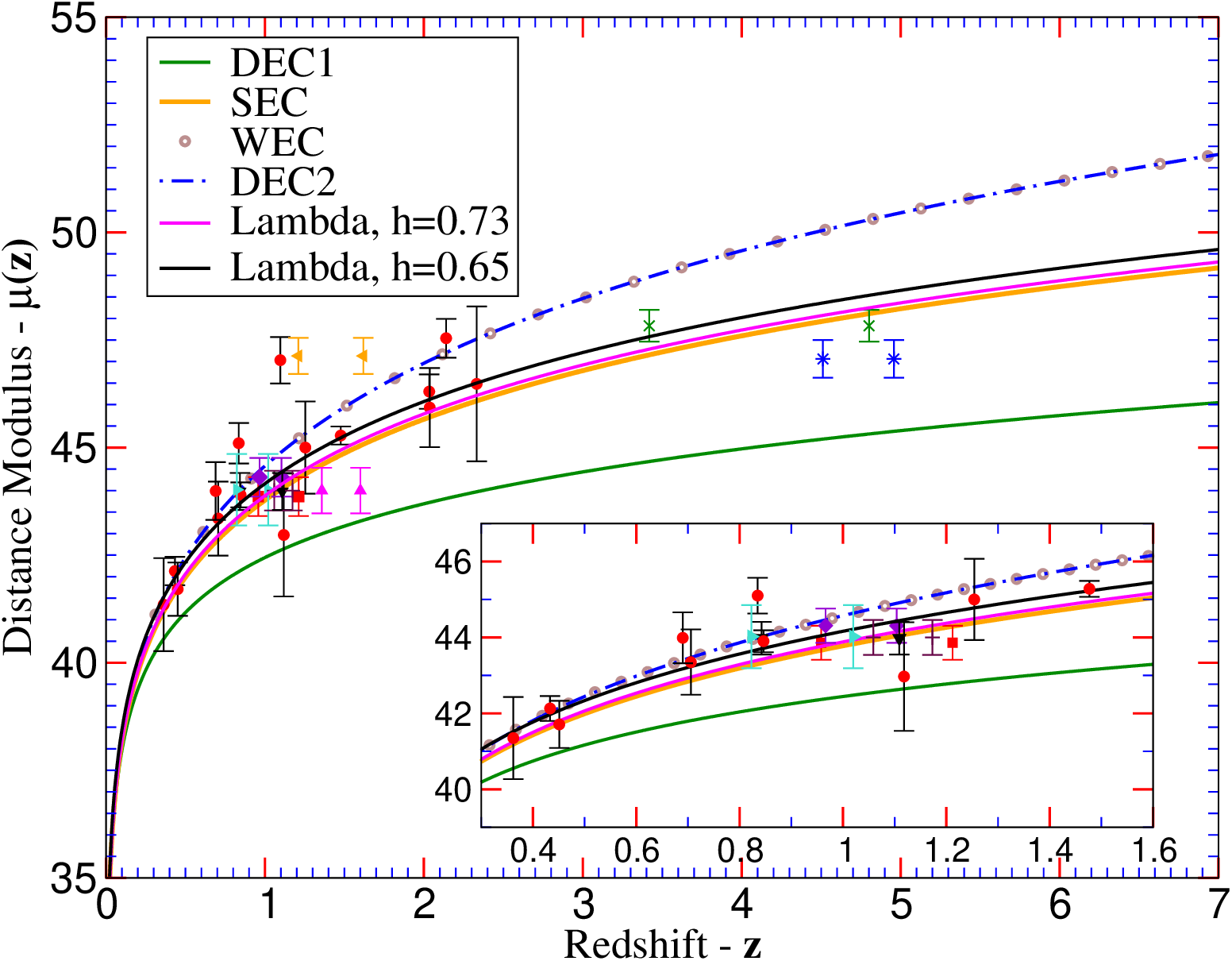},\hskip 0.15truecm \includegraphics[height=1.5in,width=1.7in]{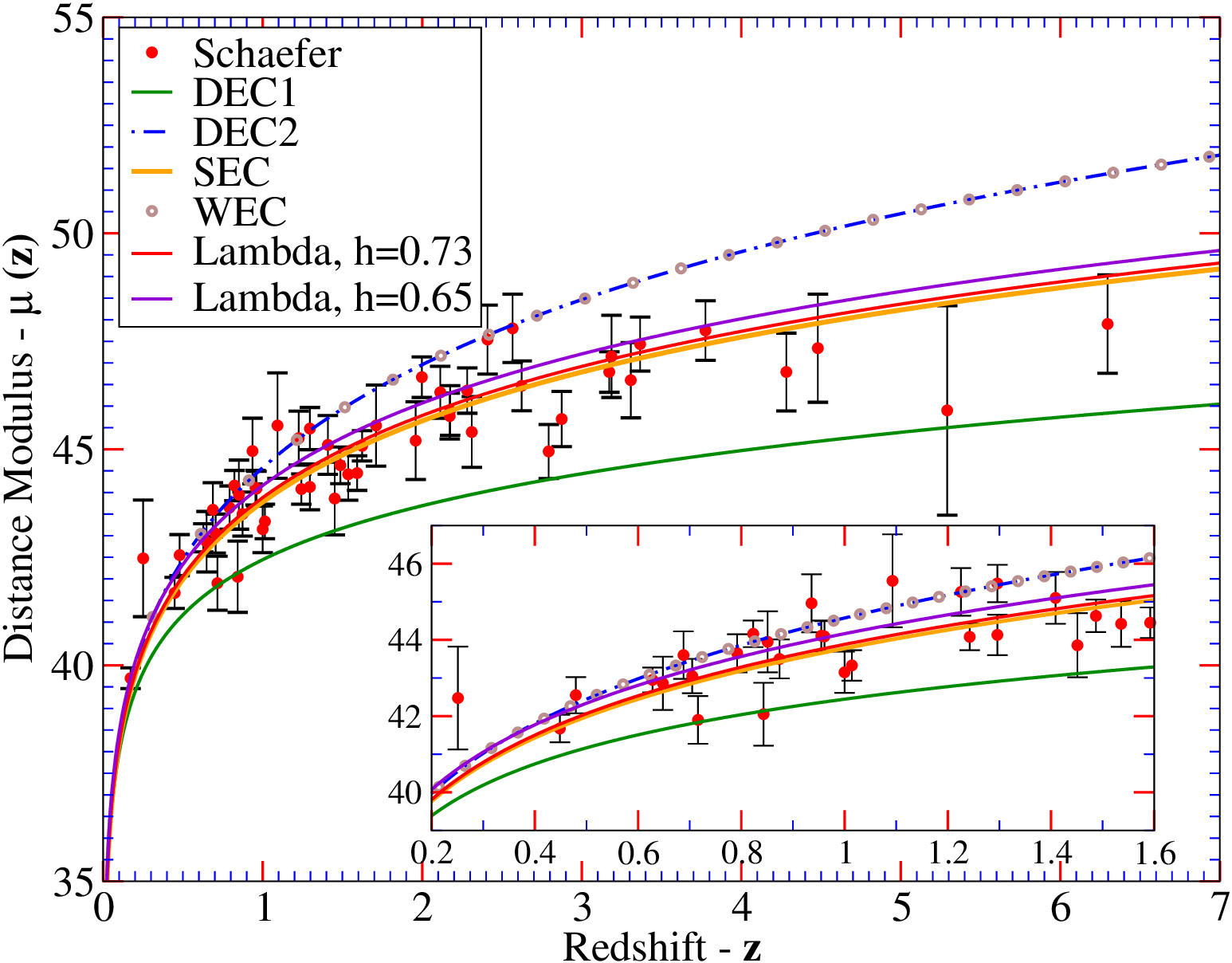},\hskip 0.15truecm \includegraphics[height=1.5in,width=1.7in]{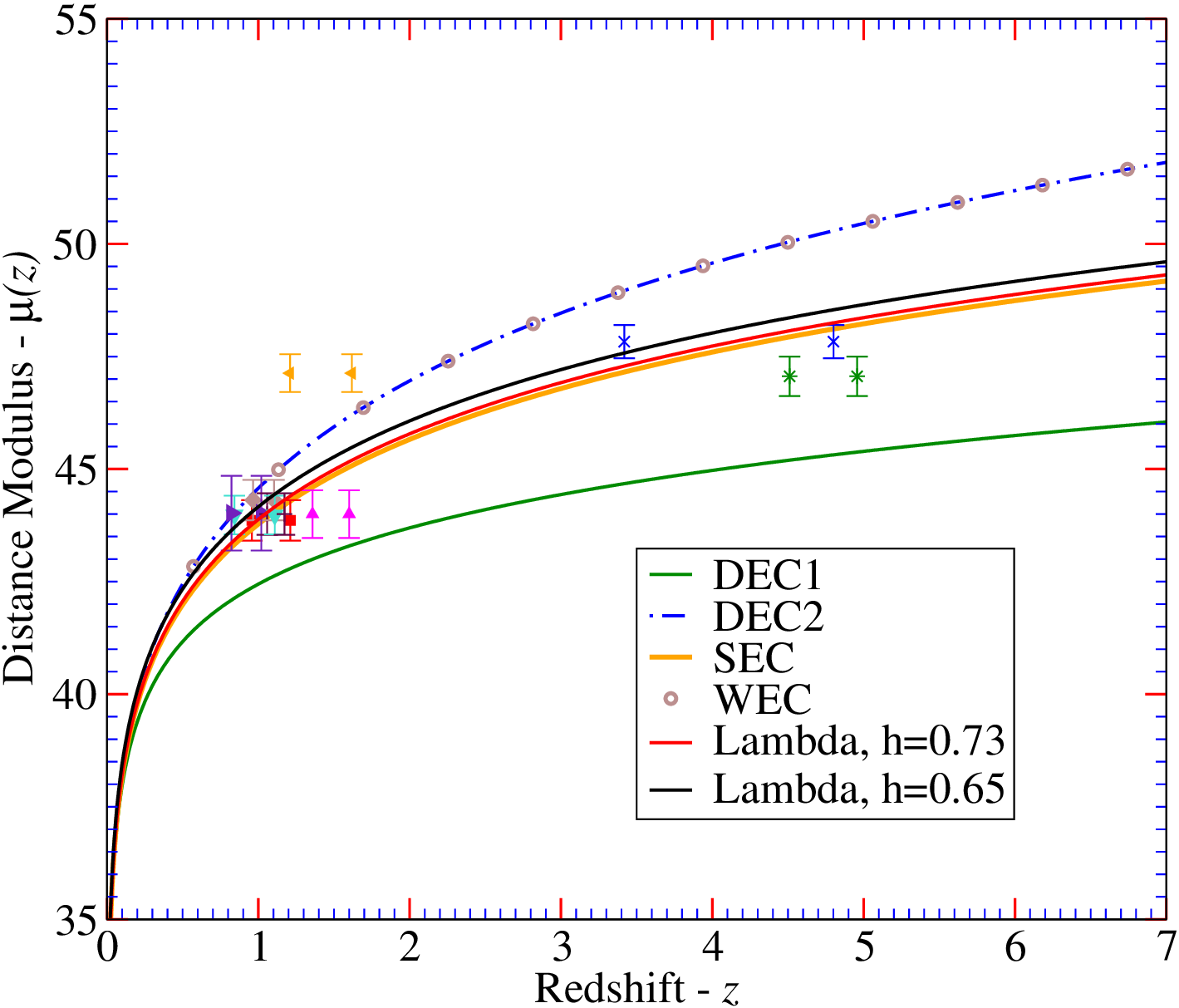},\hskip 0.15truecm \includegraphics[height=1.5in,width=1.7in]{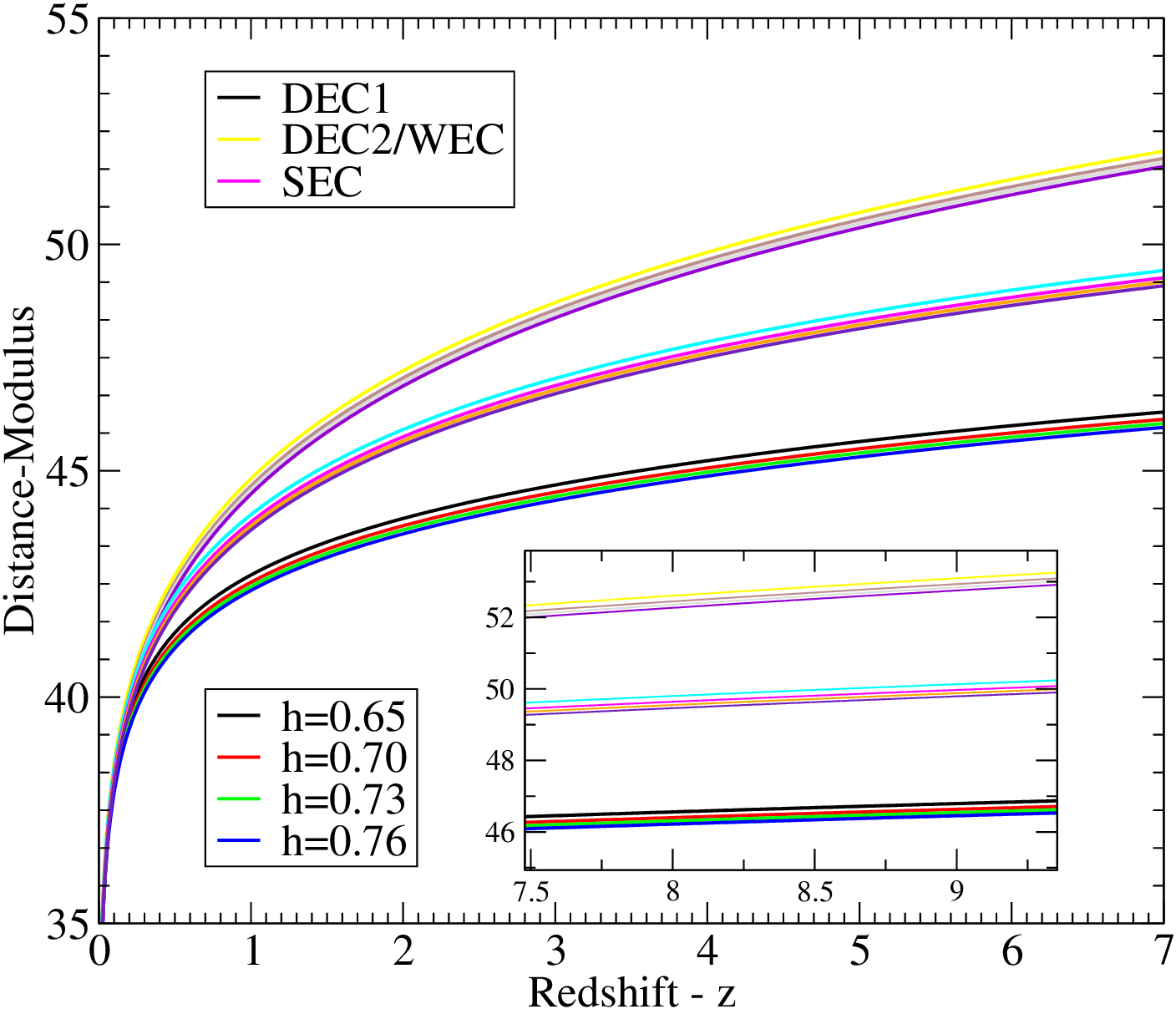}}
\caption{COLOR ONLINE - HDs for the samples used in this analysis: a) The 24 GRBs 
listed in Table-1 [Data taken from Bloom, Frail \& Kulkarni (2003)]. Notice the pairs of 
events with identical colors indicating those GRBs that had their $z$ $\theta_{\rm 
j}$-corrected with the Firmani et al. relation. b) The Schaefer's (2006) 52 GRBs events, which 
were analyzed by 5 different methods [Data taken from Schaefer presentation at the AAS Meeting 
(January 2006)]. c) The 9 GRBs listed in the last column of Table-1 that had their redshift 
corrected (shifted - the same symbol/color pairs) by using the Firmani et al. relation [Data 
taken from Firmani et al. (2006a)]. d) Behavior of the $\mu(z)$ vs. $z$ relation as a function 
of the Hubble parameter $h$ for values as indicated. No relevant changes ($\leq 1\%$) are 
observed for the interval depicted.  } 
\label{Hubble-diagram}
\end{figure*}


\section{FLRW cosmological model: Distance modulus vs. redshift relation}

Relativistic cosmology is founded on three pillars known as:

a) {\sl The Cosmological Principle}, 
which states that the universe is homogeneous and isotropic at extremely large scales. 
This leads to the FLRW line element [with $c = 1$, scale factor $R(t)$, curvature $k = 0, 
\pm 1$ and metric signature (-, +, +, +)] (D'Inverno 1993)

\be
ds^2 = - dt^2 + R^2(t) \left(\frac{dr^2}{1 - k r^2} + r^2[d\theta^2 + \sin^2\theta d\phi^2] 
\right) \; .
\label{line-element}
\ee

b) {\sl The Weyl's Postulate}, which demands that the universe's substratum be a perfect 
fluid, namely: 

\be
T_{\alpha \beta} = (\rho + p) U_\alpha U_\beta + p g_{\alpha \beta} 
\label{FLRW-line} \;,
\ee

where $U_\gamma$ is the fluid 4-velocity. 

c) {\sl The General Theory of Relativity}, that is; it should be described by Eintein's 
equations ($G = 1$, $\kappa_G = 8 \pi$)

\be
G_{\alpha \beta} - \Lambda g_{\alpha \beta} = \kappa_G T_{\alpha \beta} \; .
\label{EEqs}
\ee

Thus in a preferred coordinate frame the couple of independent (Friedmann's) equations 
reads

\begin{eqnarray}
3 \frac{\dot{R}^2 + k}{R^2} - \Lambda & = & {8 \pi } \rho \; \hskip 0.5 truecm 
 \therefore \hskip 0.5 truecm \dot{R} = \frac{dR}{dt}
\label{hubble-law},  \\
\frac{2 {R} \ddot{R} + \dot{R}^2 + k}{R^2} - \Lambda & = &  - 8\pi p \; \hskip 0.5 truecm 
\therefore \hskip 0.5 truecm \ddot{R} = \frac{d\dot{R} }{dt} \; ,
\label{friedmann-equations}
\end{eqnarray}

where $\dot{R}$ and $\ddot{R}$ are the expansion velocity and acceleration of the universe.
Putting the cosmological constant $\Lambda = 0$, from Eqs.(\ref{hubble-law}, 
\ref{friedmann-equations}) one can recast the ECs (\ref{WEC},\ref{DEC},\ref{SEC}) as 
dynamical constraints on the scale factor ($R$) and its derivatives ($\dot{R}$, $\ddot{R}$) 
whichever the curvature ($k$) come to be. The resulting set of equations reads

\begin{eqnarray}
{\rm WEC} & \Longrightarrow & - \frac{\ddot{R} }{R} + \left[\frac{\dot{R}^2 + k}{R^2} 
\right] \geq 0 \hskip 0.2 truecm  { \Longrightarrow}  \dot{R} \geq H_0 R\big\|_{k=0}, \; \; 
\forall R < R_0 ,  
\label{WEC-DYNAMIC} \\
{\rm DEC} & \Longrightarrow  &   - 2 \left(\frac{\dot{R}^2 + k}{R^2} \right) 
\leq \frac{\ddot{R}}{R} \leq \left(\frac{\dot{R}^2 + k}{R^2} \right) \label{DEC-DYNAMIC}   
\\ & \Longrightarrow & \dot{R} \leq H_0 \left[\frac{R^3_0}{R^2}\right]\big\|_{k=0}  \; \; \; 
\;{\rm and} \; \; \; \dot{R} \geq H_0 R\big\|_{k=0} , \; \; \forall R < R_0  \; , 
\nonumber \\
{\rm SEC} &  \Longrightarrow &	\frac{\ddot{R}}{R} \leq 0  
\hskip 0.3 truecm  {\Longrightarrow} \dot{R} \geq H_0 R_0, \; \; \forall R < R_0 \; .
\label{SEC-DYNAMIC}
\end{eqnarray}

Notice that physically the SEC means that in a FLRW universe the expansion is always 
decelerating whatever the curvature $k$ is. This also assures the pace with the 
attractiveness of gravity. We have also used the cosmological definition of the redshift 
$\frac{R_0}{R} = 1 + z$. Above $R_0$, and ${H}_0 = \frac{\dot{R}(t_0)}{R(t_0)}$, are 
respectively the scale factor and Hubble parameter presently (subscript $0$).


\subsection{Distance-modulus vs. redshift relation in FLRW } 

For an outlying source of apparent $m$ and absolute $M$ magnitudes, distance estimates are 
made through the {\sl distance-modulus} $m - M$, which relates to the luminosity distance 
(given in units of Gpc below)

\be
d_L = R_0 (1 + z) \int_R^{R_0} \frac{dR}{R \dot{R}} \; ,
\label{lum-distance}
\ee

through

\be
\mu({z}) \equiv m - M = 5 \log_{10} d_L ({z}) + 25 \, .
\label{distance-redshift}
\ee

By assuming a flat universe, one can obtain model-independent bounds on the 
distance-modulus $\mu(z)$ of a further-lying GRB source as a function of the 
redshift $z$ for each of the ECs enforced on an expanding FLRW universe model. 
Hence, whenever any EC gets obeyed its corresponding distance indicator should 
take values not larger than those allowed by the constraints given by 
Eqs.(\ref{WEC-DYNAMIC}, \ref{DEC-DYNAMIC}, \ref{SEC-DYNAMIC}).

\subsubsection{ECs + FLRW $\mu(z)$-$z$ relations }

After integrating the $d_L$ of Eq.(\ref{lum-distance}) for each of the ECs dynamical 
relations provided in Eqs.(\ref{WEC-DYNAMIC},\ref{DEC-DYNAMIC},\ref{SEC-DYNAMIC}) one 
obtains the distance-modulus vs. redshift relations as follows:

\begin{displaymath}
{\rm WEC + FLRW \; Hubble\; diagram} \atop \; \mu(z)-z\; \Longrightarrow  \left\{ \mu(z) 
\leq 5 \log_{10} \left[H_0^{-1} z (1 + z) \right] + 25 \right. 
\label{mu-WEC}
\end{displaymath}


\begin{displaymath}
{\rm DEC + FLRW \; Hubble\; diagram} \atop \;\mu(z)-z \; \Longrightarrow \left\{ 
\begin{array}{ll} 
& \mu(z) \geq 5 \log_{10} \left[ \frac{H_0^{-1}}{2} \frac{z (2 + z)}{1 + z} 
\right] + 25 \;\;\;\; (DEC1) \label{mu-DEC1}\\ 
& \mu(z) \leq 5 \log_{10} \left[ H_0^{-1} z (1 + z) \right] + 25  \; \;\; (DEC2) .
\label{mu-DEC2} 
\end{array} \right. 
\end{displaymath}


\begin{displaymath}
{\rm SEC + FLRW \; Hubble\; diagram} \atop \; \mu(z)-z \;  \Longrightarrow  
\left\{ \mu(z) \leq 5 \log_{10} \left[H_0^{-1} (1 + z)~ \ln(1 + z) \right] + 25 \; .
\label{mu-SEC-1} \right.
\end{displaymath}

Next we use the values $H_0 = 1/3 h$~Gpc$^{-1}$, with $h = 0.73$, to discuss the 
implications of these relations to the FLRW cosmology.

\section{Discussion and conclusions}

In Figures 1a), 1b) and 1c), we present the main results of this investigation. First, after 
looking at Figures 1a) and 1b) one can verify that the upper DEC bound (dash-dotted line)
matches perfectly the WEC $\mu(z)$ prediction (open circle line). Besides, one can also 
realize that both the DEC upper and lower bounds are respected by the GRBs observations, 
that is, each of the bursts that have been observed has its representative point in the 
FLRW HD inside the limits of DEC1 and DEC2 in Figures 1a), 1b). Besides, there 
is no clear violation of either the WEC or the SEC. This results go almost contrary to
conclusions achieved by Santos, Alcaniz \& Rebou\c cas (2006); who showed, after 
confronting the ECs with SNIa observations, that practically each of the ECs seems 
to have been violated in the recent past history of the cosmic evolution. We argue 
that as GRBs are almost free of any effect as absorption or the like, one is inclined 
to conclude that perhaps the astrophysics of SNIa has not been exhaustively understood
(Middleditch 2006), and hence the interpretation of luminosity-dimmed SNIa as messengers 
of late-time acceleration should decidely be rethought, perhaps in the lines of Middleditch 
(2006).


Further, it is quite easy to see that the SEC prediction seems to fit more properly 
the GRBs HD  than any another of the ECs+FLRW. This is quite a surprising 
result. It seems to indicate, contrarily to most of the current expectations (see Firmani 
et al. 2006b), that the cosmological constant $\Lambda$ (i.e. EoS with $\omega =-1$) is not 
the driving force pushing out the universe in the late times since it ``naturally" violates 
the SEC, as pointed out earlier. In fact, looking at Figure 1b), where Schaefer's (2006) 52 
GRBs are plotted, this behavior seems to extend down to lower redshifts $z \sim 0.1-0.5$. 
Thus our ECs+FLRW plus GRB HD seem to exclude any need for dark energy.

To bring our results into consistency, we have performed a $\chi$-square analysis to 
compare the distance-modulus vs. redshift relation as predicted by each of the ECs+FLRW 
cosmological scenarios with the HD obtained from GRBs observations. Here we 
define

\[\chi^2 = \sum_i \frac{[\mu^0_i(z) - \mu^t_i(z)]^2} {\sigma^2_{\mu^0_{i}} } \; , \] 

where respectively $ \mu^0_i(z) $ and $ \mu^t_i(z) $ define the measured and calculated 
(theoretical) value for the distance-modulus, and $\sigma_{\mu^0_{i}}$ represents the 
dispersion in the $\mu^0_i(z)$ (see Table 1). Despite the $\mu(z)$ uncertainties, the 
data clearly selects the SEC+FLRW as the best fit. Our calculations for the 24 GRBs from 
BFK2003 show that the average $\chi^2$ for the WEC 13.32, for the SEC 4.864, and for the 
DEC 26.47. For the 52 Schaefer (2006) GRBs the $\chi^2$ figures go as follows: for the 
WEC 12.17, for the SEC 1.36, and for the DEC 10.78. This therefore confirms what a visual 
inspection of Figures 1a) and 1b) allows to 
conclude. Hence, as the SEC implies that the universe is all the time decelerating 
irrespective of the type of curvature it may have, then one is forced to conclude that the 
cosmological constant is playing no role at all in the universe dynamics since it naturally 
violates the SEC. Nor it is being the fundamental field driving the purported universe 
late-time acceleration since there seems to be not such a transition down redshifts $z \sim 
0.5-1.0$. This in passing would imply that there is no need for the own dark energy field, 
as it appears to be ruled out by the GRBs HD. Besides, DE should comply with 
the ECs, and they (in particular the SEC) show that such a transition-to-acceleration at 
low redshifts seems to have not occurred.


Meanwhile, to elucidate any undisputable role of the cosmological constant we now impose 
the SEC $\rho + 3p \geq 0$ bound on the Friedmann's equations with $\Lambda \neq 0$ but 
unspecified $p=p(\rho)$ EoS\footnote{Notice that DE scenarios with $\Lambda$ constant! have 
EoS $p = \omega \rho$, with $\omega = -1$. This model clearly violates the SEC bound since 
SEC $\longrightarrow p \geq - 1/3 \rho$. However, it provides a dynamics: $\dot{R} \geq H_0 
R$, that follows the WEC bound given by the top curves of HDs in Figure 1.}, one 
obtains the dynamics (for $k=0$)

\be
\frac{\ddot{R}}{R} \leq  \frac{\Lambda}{3}  \; \Longrightarrow 
\left(\frac{\dot{R}}{R_0}\right)^2 \geq H_0^2 -  \frac{\Lambda}{3} \left[ \frac{z 
(z + 2)}{(1 + z)^2} \right] \; .
\label{lambda-neq-0}
\ee

(One can verify that for $\Lambda = 0$ one recovers Eq.(\ref{SEC-DYNAMIC}), above).
Thus, the luminosity-distance becomes

\be
d_L \leq  \frac{(1 + z)}{\left(H_0^2 - \frac{\Lambda}{3} \left[ \frac{z 
(z + 2)}{(1 + z)^2} \right] \right)^{1/2}} \ln(1 + z) \; ,
\label{dl-lambda}
\ee

and therefore, the $\mu(z)$-$z$ HD reads:

\be
{\rm SEC + FLRW + \Lambda \; Hubble\; diagram} \atop \; \mu(z)-z \;  \Longrightarrow  
\left\{ \mu(z) \leq 5 \log_{10} \left( \frac{(1 + z) \; \ln(1 + z) }{\left(H_0^2 - 
\frac{\Lambda}{3} \left[ \frac{z (z + 2)}{(1 + z)^2} \right] \right)^{1/2}} \right) 
+ 25 
\label{mu-SEC} \right. .
\ee

Putting $\Lambda = 2.036 \times 10^{-35}$ s$^{-2} = 0.214 {\rm Gpc}^{-2}$, and $h=0.73$ 
one verifies that according to Figures 1a), 1b) and 1c) it gets close to the SEC limit of 
Eqs.(\ref{hubble-law},\ref{friedmann-equations}) without $\Lambda$. The $\chi^2$ analysis 
for this case gives 1.33 for the 52 GRBs and 4.36 for the 24 GRBs. It is a bit better than 
the SEC limit alone. However, the error bars do not allow to claim that there is any actual 
difference between the $\chi^2$ value for SEC+FLRW+$\Lambda$ and the sole for SEC+FLRW. Thus, 
this result can be interpreted in the lines of our discussion above as either indicating that 
there is no need for $\Lambda$, or that $\Lambda$ is time-variable. As the HD for the case 
$\Lambda \neq 0$ plus SEC+FLRW does not coincide with the one corresponding to the case $p= 
-\rho$ (footnote {8}) then one is led to conclude that the cosmo-(i)logical constant does not 
exist at all or it is not constant either. Figures 1 illustrate the behavior of a 
$\Lambda$+SEC+FLRW for different values of the Hubble parameter $h$.


In summary, despite the relatively large uncertainties still present in the luminosity 
distance of most GRBs and the small statistics of GRB with low redshifts, the approach 
introduced here appears to be more conclusive than 
an analysis standing on SNIa data alone. The main reason is that the sample of GRBs 
that we are using for the present analysis already reaches farther out, upto redshifts 
$z\sim 7$ where the early-days cosmology was playing hard, while SNIa are naturally limited 
to distance scales no farther out than $z \sim 1$. (Surely, the set of both lower (about
$z \sim 0.1-0.5$), and larger (upto $z \sim 10-20$) redshifts will raise in the near future 
when more GRBs are to be observed by the SWIFT 
satellite).  On physical grounds, the outcome of this inedit test suggests that the FLRW 
model when faced with the GRB HD guarantees the prevalence of the WEC, SEC and 
DEC bounds, with the SEC being the better cosmological fit to the GRBs observational data.
As SNIa are facing problems to go a pace with the ECs+FLRW, we think that it is the SNIa 
astrophysics that should be reviewed instead of general relativity (Middleditch 2006).

To the last, the GRBs HD has been built to demonstrate the reliability of 
GRBs as an independent, but nonetheless complementary tool to SNIa (and perhaps to CMB 
observations), to confront current cosmological models. Thus, everything now allows to 
forsee a promising future for a high-precision era in GRBs cosmology, a field which is 
by-now at its infancy (Lazzati et al. 2006), based on these novel relations that promise 
to become a major breakthrough in GRBs cosmology. This is the first paper of a thorough 
research of this key issue in current Cosmology. In forthcoming papers other cosmological 
models, specially those purporting modifications of gravity at very larger scales, will be 
analyzed in face of GRBs data to cross-check them for consistency with the HD of GRBs.  



\begin{thebibliography}{200}


\bibitem[Amati et al.(2002)]{amati02} Amati L. et al., 2002, A\&A,  390, 81


\bibitem[Blain \& Natarajan (2000)]{BN00} Blain A. W. \& Natarajan P., 
2000, MNRAS, 312, L35

\bibitem[Bloom, Frail \& Kulkarni(2003)]{Bloom03} Bloom J. S., Frail D.~A.
\& Kulkarni S.~R., 2003, ApJ, 594, 674

\bibitem[Bromm \& Loeb (2002)]{BL02} Bromm V. \& Loeb A., 2002, ApJ, 
575, 111

\bibitem[Costa et al.(1997)]{Costa97} Costa E. et al., 1997, Nature, 
387, 783

\bibitem[Dai, Liang \& Xu(2004)]{Dai04} Dai Z.G., Liang E.W. \& Xu D.,
2004, ApJ, 612, L101

\bibitem{dinverno}R. D'Inverno, Introducing Einstein's relativity, 
(Clarendoss Press, Oxford) (1993)


\bibitem[Firmani et al.(2004)]{Firmani04} Firmani C., Avila-Reese V.,
Ghisellini G. \& Tutukov A. V., 2004, ApJ, 611, 1033

\bibitem[Firmani et al.(2005)]{Firmani05} Firmani C., Ghisellini G.,
Ghirlanda G. \& Avila-Reese V., 2005, MNRAS, 360, L1

\bibitem[Firmani et al.(2006a)]{Firmani06a} Firmani C., Ghisellini G.,
Avila-Reese V. \& Ghirlanda G., 2006a, MNRAS, 370, 185

\bibitem[Firmani et al.(2006b)]{Firmani06b} Firmani C.,  Avila-Reese V.,
Ghisellini G., \& Ghirlanda G., 2006b, astro-ph/0605267

\bibitem[Frail et al.(1997)]{Frail97} Frail D.~A. et al., 1997, Nature, 
389, 261

\bibitem[Friedman \& Bloom (2005)]{FB05} Friedman A. S. \& Bloom J. S., 
2005, ApJ, 627, 1

\bibitem[Friedman \& Bloom (2006)]{FB06} Friedman A. S., Graduate Research 
Forum, March 14 (2006): {\sl The promise and limitations of gamma-ray burst 
standard candles} 

\bibitem{ghirlanda04}Ghirlanda G., Ghisellini G. \& Lazzati D., 2004;

\bibitem{hubble1929}Hubble, E. P., 1929, Proc. Nat. Acad. Sci. USA, 15, 168-173

\bibitem[Lamb \& Reichart(2000)]{Lamb00} Lamb D. Q, \& Reichart D. E., 
2000, ApJ, 536, 1

\bibitem{lazzati2004} D. Lazzati et al., (2006), astro-ph/0602216

\bibitem[Liang \& Zhang(2005)]{LZ05} Liang  E. \& Zhang B., 2005, ApJ, 
633, 611,  (LZ2005)

\bibitem[Lloyd-Ronning, Fryer \& Ramirez-Ruiz(2002)]{LFR02} 
Lloyd-Ronning N. M., Fryer C. L. \& Ramirez-Ruiz E., 2002, ApJ, 574, 554

\bibitem[Lloyd-Ronning \& Ramirez-Ruiz (2002a)]{Lloyd02a} Lloyd-Ronning N. 
\& Ramirez-Ruiz E., 2002, ApJ, 576, 101

\bibitem{middledicth06} J. Middledith, astro-ph/0608386 (2006)

\bibitem{sollerman}E. M\"ortsell \& J. Sollerman, (2005), JCAP 06, 009


\bibitem[Paczynski (1998)]{Paczynski98} Paczynski B., 1998, ApJ, 494, L45

\bibitem{santiago} S. E. P\'erez Bergliaffa, astro-ph/0608072 (2006)


\bibitem{b.b}Santos J., Alcaniz  J. \& Rebou\c cas M., report astro-ph/0608072 (2006)

\bibitem[Schaefer(2003)]{Schaefer03} Schaefer B.E., 2003, ApJ, 583, L67

\bibitem[Schaefer(2006)]{Schaefer06}Schaefer B.E., 2006, paper presented 
at the AAS Meeting January (2006)

\bibitem[Totani(1997)]{Totani97} Totani T., 1997, ApJ, 486, L71

\bibitem[van Paradijs et al. (1997)]{vParadijs97} van Paradijs J. et 
al., 1997, Nature, 386, 686

\bibitem{visser96} M. Visser, {\sl Lorentzian wormholes}, AIP (1996)
(Woodbury, New York)

\bibitem[Wijers et al. (1998)]{Wijers98} Wijers R. A. M. J., Bloom 
J. S., Bagla J. S. \& Natarajan P., 1998, MNRAS, 249, L13

\bibitem[woosley2006]{woosley-bloom06}S. E. Woosley \& J. S. Bloom, 2006, 
Annu. Rev. Astron. Astrophys. 44, 507-556  

\bibitem[Xu (2005)]{Xu05} Xu D., 2005, preprint (astro-ph/0504052)

\bibitem[Yonetoku et al.(2004)]{Yonetoku04} Yonetoku D. et al., 2004, 
ApJ, 609, 935


\end{thebibliography}
\end{document}